\documentclass[preprint,pra,showpacs,nofootinbib]{revtex4-1}  

\usepackage{amsmath,amssymb,amsthm}
\usepackage[mathscr]{euscript}
\usepackage{graphicx}  
\usepackage{float}
\usepackage{dcolumn}   
\usepackage{bm}        
\usepackage{subfig}    
\usepackage{hyperref}  

\newcommand{\bea}{\begin{eqnarray}}
\newcommand{\eea}{\end{eqnarray}}
\newcommand{\beq}{\begin{equation}}
\newcommand{\eeq}{\end{equation}}

\begin{document}

\title{Position-dependent thermalization of two-level probes in de Sitter spacetime: Interplay between Gibbons-Hawking and Unruh effects}
\author{  Yao Jin\footnote{Corresponding author. yaojin@gyu.edu.cn} }
\affiliation{School of Science,
Guiyang University,  Guiyang, Guizhou 550005, China}


\begin{abstract}
We investigate the thermalization process of two-level probes in de Sitter spacetime from a quantum-metrological perspective. For a static probe separated by a finite distance from a freely falling observer, both the intrinsic Gibbons--Hawking temperature and the position-dependent Unruh temperature are encoded in the probe state. The mutual influence between the encoding rate of the Gibbons--Hawking temperature and that of the Unruh temperature is studied. Results show that the thermalization rate is equal to the sum of the encoding rates of the Gibbons--Hawking temperature and that of the Unruh temperature. Interestingly, a nonzero inherent acceleration induced by the separation between the probe and a freely falling observer does not necessarily suppress the encoding of the Gibbons--Hawking temperature. Instead, there exists an optimal inherent acceleration at which the encoding rate of the Gibbons--Hawking temperature is maximized. Furthermore, the required total probe time of estimating the thermal effect with sufficient precision is shown in relation with the thermalization rate. The required total probe time and the corresponding number of probes remain experimentally feasible when the position-dependent inherent acceleration satisfies $\frac{a}{2\pi\omega_0}\rightarrow0.1$.
\end{abstract}

\maketitle

\section{Introduction}

De Sitter spacetime, which possesses the same degree of symmetry as Minkowski spacetime, has attracted considerable attention over the past decades. It plays a central role in modern cosmology. According to current observations and inflationary theories, our universe may approach de Sitter geometry in both the far past and the far future. Moreover, a holographic duality may exist between quantum gravity in de Sitter spacetime and a conformal field theory living on its timelike infinity boundary~\cite{Strominger01a,Strominger01b}.
In a seminal work, Gibbons and Hawking showed that the ratio of the probabilities for a freely falling detector to absorb and emit a particle with energy $E$ is given by $e^{-2\pi E\sqrt{3/\Lambda}}$, where $\Lambda$ is the cosmological constant. This result indicates that the detector perceives an isotropic thermal radiation background characterized by the Gibbons--Hawking temperature $T_{\mathrm{GH}} = \frac{1}{\pi}\sqrt{\frac{\Lambda}{12}}$~\cite{Gibbons77}. Since then, this thermal character has been investigated through various approaches, including the embedding of four-dimensional de Sitter spacetime into a five-dimensional flat spacetime~\cite{Deser97}, the thermalization of detectors within the framework of open quantum systems~\cite{Yu11}, and a variety of related physical effects~\cite{Narnhofere96,Hutasoit09,Henry10,Zhu08,Zhou10,Tian13,Oshita14,Zhou}. These studies include investigations of the spontaneous excitation rate~\cite{Zhu08}, the Lamb shift~\cite{Zhou10}, the geometric phase~\cite{Tian13} of static and freely falling detectors, the Brownian motion of a particle coupled to vacuum fluctuations in de Sitter spacetime~\cite{Oshita14}, and the interaction potential between two static detectors~\cite{Zhou}.
The effective temperature perceived by a detector depends on its trajectory. For a pointlike freely falling detector, the detector thermalizes as if it were immersed in a thermal bath at the Gibbons--Hawking temperature $T_{\mathrm{GH}}$. For a static detector, the effective temperature is given by $T_S = \sqrt{T_{\mathrm{GH}}^2 + T_U^2}$, where $T_U$ is the Unruh temperature associated with the detector's inherent acceleration.
During the thermalization process, both the intrinsic Gibbons--Hawking temperature and the position-dependent Unruh temperature are encoded in the state of the detector. This naturally raises two questions: Does the encoding of one temperature affect the encoding of the other? And how can the temperature encoding rate be quantified?

To address the latter question, we first recall the metrological procedure for parameter estimation. To estimate an unknown parameter encoded in a quantum state, a large number of measurements are generally required. Measurements are performed on individual probes, and their outcomes follow a probability distribution determined by the parameter of interest. By constructing an appropriate estimator, the parameter can be inferred with an uncertainty bounded by the Cram\'{e}r--Rao inequality~\cite{Helstrom,Holevo,Hubner,Braunstein}. The attainable precision is governed by the Fisher information, which quantifies the information about the parameter extracted from measurements and determines the ultimate sensitivity of parameter estimation.
Quantum metrology has been successfully applied to a wide range of precision-measurement tasks, including quantum frequency standards~\cite{Bollinger}, optimal quantum clocks~\cite{Buzek}, measurements of gravitational acceleration~\cite{Peters}, and clock synchronization~\cite{Jozsa}.

In the thermalization process considered here, a fixed total probe time can be distributed among a large number of probes. By optimizing the evolution time, initial state, and measurement basis of each probe, the precision of temperature estimation can be optimized under this fixed resource constraint. This motivates the introduction of a temperature encoding rate, which characterizes the amount of temperature information that can be extracted per unit total probe time. Such a metrological description provides a quantitative way to investigate how the different thermal contributions perceived by a detector are encoded in its quantum state.
The paper is organized as follows. In Sec.~II, we review the evolution of a two-level probe atom coupled to a fluctuating scalar field in de Sitter spacetime. In Sec.~III, we introduce the temperature encoding rate from a quantum-metrological perspective under a fixed total probe time. In Sec.~IV, we analyze the encoding rates of the effective, intrinsic, and Unruh temperatures. In Sec.~V, we analyze the minimum total probe time required to estimate the corresponding thermal effects. Finally, we summarize our results in Sec.~VI.

\section{Evolution of the probe two-level atom coupled to fluctuating scalar field in de sitter space-time}
We begin by review the evolution of a two-level atom coupled to fluctuating scalar field. In the present work, we use the natural units with $\hbar=c=k_B=1$.
The total Hamiltonian of a two-level atom and the scalar field has the form
$
H=H_s+H_f+H_I
$.
Here $H_s={1\over
2}\,\omega_0\sigma_3$ denotes the Hamiltonian of
the two-level atom with $\omega_0$ denoting the transition frequency and $\sigma_l$ denoting the $l$th Pauli matrix. $H_f$ denotes the
Hamiltonian of the free scalar field. $H_I$ denotes the the interaction Hamiltonian
between the two-level atom and the scalar field with the form $H_I(\tau)=\mu(e^{i\omega_0\tau}\sigma_++e^{-i\omega_0\tau}\sigma_-)\phi(x(\tau))$. Here $\sigma_+$ and $\sigma_-$ denote the atomic
rasing and lowering operator, $\mu$ denotes coupling costant,
$\phi(x)$ denotes the scalar field strength.

We assume the scalar field is initially in the de Sitter invariant vacuum~\cite{Allen85,Polarski89,Galtsov}. So the initial total density matrix $\rho_{tot}(0)=\rho(0) \otimes |0\rangle\langle0|$ with $\rho(0)$ denoting the initial reduced density matrix of the atom, and $|0\rangle$ denoting the de Sitter invariant vacuum of the scalar field. The evolution of the total density matrix $\rho_{tot}$ at time $\tau$ satisfies
\begin{equation}\label{evo}
\frac{\partial\rho_{tot}(\tau)}{\partial\tau}=-i[H,\rho_{tot}(\tau)]\;.
\end{equation}
In the weak coupling assumption, the evolution of the reduced
density matrix $\rho(\tau)$ is given in the
Kossakowski-Lindblad form~\cite{Lindblad, pr5} as
\begin{equation}\label{master}
{\partial\rho(\tau)\over \partial \tau}= -i\big[H_{\rm eff},\,
\rho(\tau)\big]
 + {\cal L}[\rho(\tau)]\ ,
\end{equation}
where
\begin{equation}
{\cal L}[\rho]={1\over2} \sum_{i,j=1}^3
a_{ij}\big[2\,\sigma_j\rho\,\sigma_i-\sigma_i\sigma_j\, \rho
-\rho\,\sigma_i\sigma_j\big]\ .
\end{equation}
The above coefficients of the Kossakowski matrix $a_{ij}$ are expressed as
$
a_{ij}=A\delta_{ij}-iB
\epsilon_{ijk}\delta_{k3}-A\delta_{i3}\delta_{j3}\;,
$
with
$
A=\frac{1}{4}[{\cal {G}}(\omega_0)+{\cal{G}}(-\omega_0)]\;,\;~~
B=\frac{1}{4}[{\cal {G}}(\omega_0)-{\cal{G}}(-\omega_0)]\;.
$
Here
$
{\cal G}(\lambda)=\int_{-\infty}^{\infty} d\Delta\tau \,
e^{i{\lambda}\Delta\tau}\, G^{+}\big(\Delta\tau\big)
\;,
$
with $G^{+}(x-x')$ being shown in relation with the correlation function of the scalar field $\langle0|\phi(x)\phi(x')|0 \rangle$ as
$
G^{+}(x-x')=\mu^2\,\langle0|\phi(x)\phi(x')|0 \rangle\;.
$
Absorbing the Lamb shift term, the effective Hamiltonian $H_{\rm eff}$ is written as
$
H_{\rm eff}=\frac{1}{2}\Omega\sigma_3={1\over 2}\{\omega_0+{i\/2}[{\cal
K}(-\omega_0)-{\cal K}(\omega_0)]\}\,\sigma_3\;,
$
with $\Omega$ denoting the effective level spacing of the atom, and
$
{\cal K}(\lambda)=\frac{P}{\pi
i}\int_{-\infty}^{\infty} d\omega\ \frac{ {\cal G}(\omega)
}{\omega-\lambda}\;.
$
The initial state of the two-level atom in each probe is assumed to be prepared in pure state as $\cos\frac{\theta}{2}|+\rangle+\sin\frac{\theta}{2}e^{i\phi}|-\rangle$ with $\theta$, and $\phi$ denoting the weight and phase factor, $|+\rangle$ and $|-\rangle$ denoting the excited and ground state of the atom.
The reduced density matrix can be expanded as $\rho=\frac{1}{2}(I+\mathbf{\omega}\cdot\mathbf{\sigma})$ with $I$ denoting the unit matrix, $\mathbf{\sigma}$ denoting the Pauli matrix and $\mathbf{\omega}$ denoting the Bloch vector. Therefore, the Bloch vector with proper time $\tau$ is calculated as:
\begin{eqnarray}\label{Bloch}
&&\omega_1(\tau)=\sin\theta\cos(\Omega\tau+\phi)\,e^{-2A\tau}\;,\nonumber\\
&&\omega_2(\tau)=\sin\theta\sin(\Omega\tau+\phi)\,e^{-2A\tau}\;,\\
&&\omega_3(\tau)=\cos\theta\, e^{-4A\tau}-\frac{B}{A}(1-e^{-4A\tau})\nonumber\;.
\end{eqnarray}
The atomic evolution factors $A$ and $B$ are determined by the field correlation function, while the field correlation function is determined by the space-time itself.
De Sitter spacetime can be represented as the hyperboloid
\begin{align}
z_0^2 - z_1^2 - z_2^2 - z_3^2 - z_4^2 = -\alpha^2,
\end{align}
embedded in five-dimensional Minkowski space with the metric
\begin{align}
ds^2 = dz_0^2 - dz_1^2 - dz_2^2 - dz_3^2 - dz_4^2 .
\end{align}
Here $\alpha = \sqrt{3/\Lambda}$.
Through the parametrization
\begin{align}
z_0 &= \sqrt{\alpha^2 - r^2}\sinh(t/\alpha), \nonumber\\
z_1 &= \sqrt{\alpha^2 - r^2}\cosh(t/\alpha), \nonumber\\
z_2 &= r\cos\theta, \nonumber\\
z_3 &= r\sin\theta\cos\varphi, \\
z_4 &= r\sin\theta\sin\varphi, \nonumber
\end{align}
the static de Sitter metric is obtained as
\begin{align}
ds^2 = \left(1 - \frac{r^2}{\alpha^2}\right) dt^2
      - \left(1 - \frac{r^2}{\alpha^2}\right)^{-1} dr^2
      - r^2 d\theta^2
      - r^2 \sin^2\theta\, d\varphi^2 .
\label{de-Sitter-metric}
\end{align}
The sphere $r=\alpha$ is singular, and it corresponds to the cosmological horizon. A static atom has the inherent acceleration
\begin{eqnarray}\label{proper acceleraion}
a=\frac{r}{\alpha^2}\big(1-\frac{r^2}{\alpha^2}\big)^{-1/2}.
\end{eqnarray}

The field correlation function has the form
\begin{align}
\langle 0|\phi(x)\phi(x')|0\rangle=-\frac{1}{4\pi^2}\Big[(z_0-z'_0-i\epsilon)^2-\sum_{i=1}^4(z_i-z'_i)^2\Big]^{-1}\;.
\end{align}
Applying the trajectory of a static atom, we have
\begin{eqnarray}\label{wightman function 3}
\langle 0|\phi(x)\phi(x')|0\rangle=-\frac{1}{16\pi^2\kappa^2\sinh(\frac{\tau-\tau'}{2\kappa}-i\epsilon)},
\end{eqnarray}
where $\kappa=\sqrt{g_{00}}\alpha$ and $\tau=\sqrt{g_{00}}t$.
Here we note we let $\varepsilon=0$ after the calculation.
All coefficients that affect the evolution of the atom are calculated as
$
A=\frac{1}{4}\gamma_0\gamma_r\;,
$
$
B=\frac{1}{4}\gamma_0\;,
$
with
\begin{equation}
\gamma_r=\frac{e^{1/X_{eff}}+1}{e^{1/X_{eff}}-1}\;,
\end{equation}
and
$
\Omega=\omega_0+\frac{\gamma_0}{2\pi\omega_0^3}\,P\int_0^\infty
d\omega\,\omega^3\left(\frac{1}{\omega+\omega_0}-\frac{1}{\omega-\omega_0}\right)\left(1+\frac{2}{e^{1/X_{eff}}-1}\right)\,.
$
Here $\gamma_0=\frac{\mu^2\omega_0}{\pi}$ denotes the spontaneous emission rate in Minkowski-vacuum, and $X_{eff}=\frac{1}{2\pi\kappa\omega_0}$ denotes the dimensionless effective temperature. For inertial observer with $r=0$, such factor approaches to $X_\alpha=\frac{1}{2\pi\alpha\omega_0}$, which only depends on the cosmological constant, and is dubbed as the dimensionless intrinsic temperature. We let $X_a=\frac{a}{2\pi\omega_0}$ denote the dimensionless Unruh temperature. Therefore, the dimensionless effective temperature can be expressed through the dimensionless intrinsic temperature and the dimensionless Unruh temperature as
\begin{equation}
X_{eff}^2=X_\alpha^2+X_a^2\;.
\end{equation}
Therefore, the Bloach vector of atomic state can be written as
\begin{eqnarray}
&&\omega_1(\tau)=\sin\theta\cos(\Omega\tau+\phi)\,e^{-\gamma_r\gamma_0\tau/2}\;,\nonumber\\
&&\omega_2(\tau)=\sin\theta\sin(\Omega\tau+\phi)\,e^{-\gamma_r\gamma_0\tau/2}\;,\\
&&\omega_3(\tau)=\cos\theta\, e^{-\gamma_r\gamma_0\tau}-\frac{1}{\gamma_r}(1-e^{-\gamma_r\gamma_0\tau})\nonumber\;.
\end{eqnarray}

Apparently, the distribution of field modes affects the atomic evolution coefficient $\gamma_r$ that $\gamma_r$ is determined by the cosmological constant $\Lambda$ and the separation distance $r$ from a free falling observer.
\section{Quantifying the temperature encoding rate from metrological process under a fixed total probe time}

To estimate the $\Lambda$-dependent intrinsic temperature as well as the position $r$-dependent Unruh temperature, large number $N$ of probes are needed. In each probe, the probe two-level atom is prepared in proper initial state, and evolves with the fluctuation scalar field with proper evolution time $\tau$, and then is measured through proper measurement in basis of
$\{\cos[\frac{\theta'}{2}+(-1)^j\frac{\pi}{2}]|+\rangle+e^{i\phi'}\sin[\frac{\theta'}{2}+(-1)^j\frac{\pi}{2}]|-\rangle\}$, $(j=0,1)$.
Here $\theta'$ and $\phi'$ are arbitrary weight and phase factor of the basis of measurement, and $j$ denotes the result of measurement.
Therefore, the probability of result $j$ becomes
\begin{equation}
P_j=\frac{1}{2}[1+(-1)^j\omega_1(\tau)\sin\theta'\cos\phi'+(-1)^j\omega_2(\tau)\sin\theta'\sin\phi'+(-1)^j\omega_3(\tau)\cos\theta']\;.
\end{equation}
Following the measurements results of $N$ probes, using maximum-likelihood estimation, arbitrary parameter $X$ can be estimated with the uncertainty of the estimation satisfying~\cite{Helstrom,Holevo,Hubner,Braunstein}
$
Var (X)\geq\frac{1}{N F_X}
$
with~\cite{J1,J2}
\begin{equation}\label{F}
F_X=\sum_j\frac{(\partial_X P_j)^2}{P_j}=\frac{[\partial_X \omega_1(\tau)\sin\theta'\cos\phi'+\partial_X \omega_2(\tau)\sin\theta'\sin\phi'+\partial_X \omega_3(\tau)\cos\theta']^2}{1-[\omega_1(\tau)\sin\theta'\cos\phi'+\omega_2(\tau)\sin\theta'\sin\phi'+\omega_3(\tau)\cos\theta']^2}\;.
\end{equation}
We use $T$ to denote the total time of $N$ probes, so $N=\frac{T}{\tau}$, and the uncertainty of the estimation of parameter $X$ satisfies
\begin{equation}
Var (X)\geq\frac{1}{N F_X}=\frac{1}{T\gamma_0 R_X}\;,
\end{equation}
with
\begin{equation}
R_X=\frac{F_X(\rho(0),\gamma_0\tau)}{\gamma_0 \tau}\;.
\end{equation}
Therefore, the precision of estimation of parameter $X$ in a fixed total probe time $T$ is determined by the factor $R_X$. The larger of $R_X$ becomes, the less uncertainty of $X$ that we can obtain in a fixed total probe time $T$. So we use $R_X$ to denote the encoding rate of parameter $X$. The thermalization rate is quantified through the encoding rate of the effective temperature $X_{eff}$. 

\begin{figure}[htbp]
\centering
\subfloat[$X_{eff}=0.8$]{%
    \includegraphics[height=2.1in, keepaspectratio]{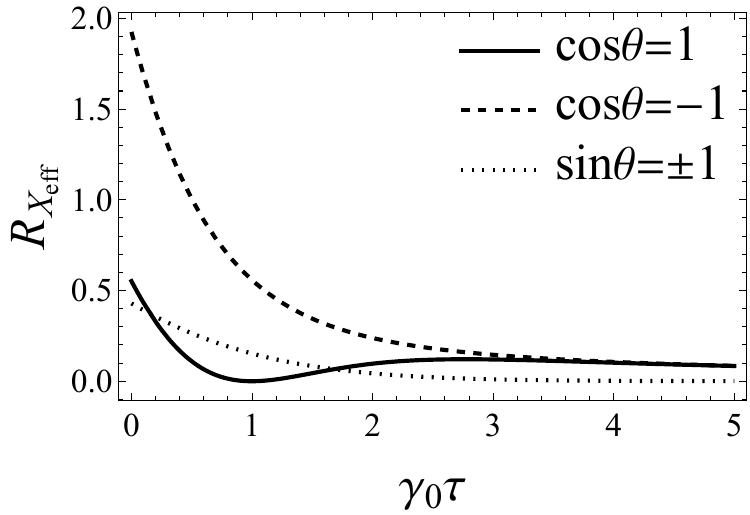}%
}%
\hfill
\subfloat[$X_{eff}=5$]{%
    \includegraphics[height=2.1in, keepaspectratio]{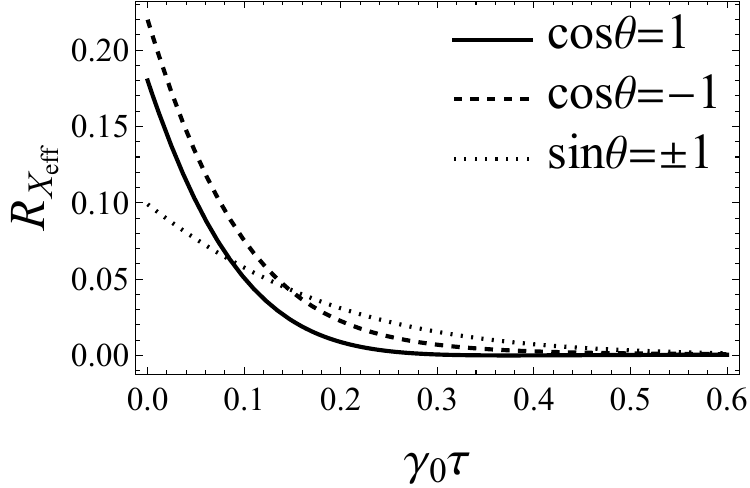}%
}%
\caption{$R_{X_{eff}}$ as a function of $\gamma_0\tau$ for initial states $\cos\theta=1$, $\cos\theta=-1$, and $\sin\theta=\pm1$, respectively, at $X_{eff}=0.8$ and $5$.}
\label{F4}
\end{figure}

Largest temperature encoding rate $R_{X_{eff}}$ is obtained after optimizing the evolution time $\tau$, the initial probe state $\rho(0)$ and its corresponding measurement basis in each probe. The optimal measurement basis in each probe is determined by the initial state of the probe atom.
For initial state $\frac{1}{\sqrt{2}}(|+\rangle\pm e^{i\phi}|-\rangle)$ with $\sin\theta=\pm1$, the largest $R_{X_{eff}}$ is reached by choosing measurement basis with $|\sin\theta'|=1$, $\phi'=\Omega\tau+\phi$.
While for initial state $|+\rangle$ and $|-\rangle$ with $\cos\theta=1$ and $\cos\theta=-1$, the largest $R_{X_{eff}}$ is obtained by choosing measurement basis with $|\cos\theta'|=1$.
To optimize the initial state and the evolution time in each probe, we plot $R_{X_{eff}}$ in function of $\gamma_0\tau$ in small temperature case with $X_{eff}=0.8$ and large temperature case with $X_{eff}=5$ in Fig.~(\ref{F4}). The solid line, dashed line, dotted line denote the cases with initial excited state $\cos\theta=1$, initial ground state $\cos\theta=-1$, and initial equal weight superposition state $\sin\theta=\pm1$ respectively.
Results show that initial ground state always show advantages, and the corresponding optimal evolution time becomes $\gamma_0\tau\rightarrow0$. In the following discussions, we use such settings.
\section{The effects of dimensionless Unruh temperature on the encoding rate of dimensionless intrinsic temperature}
\begin{figure}[htbp]
\centering
\includegraphics{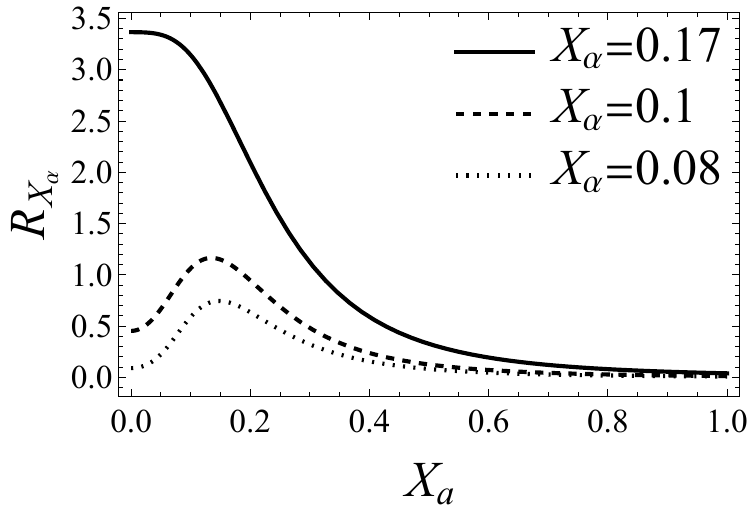}
\caption{ $R_{X_{\alpha}}$ as function of $X_a$ with intrinsic temperature $X_{\alpha}=0.17, 0.1, 0.08$ respectively.
}\label{F2}
\end{figure}

As is shown before, the dimensionless effective temperature $X_{eff}$ is determined by the dimensionless intrinsic temperature $X_{\alpha}$ and the dimensionless Unruh temperature $X_a$. The encoding rates of the temperatures can be written as
$
R_{X_{eff}}=R_{\gamma_r}(\frac{\partial\gamma_r}{\partial X})^2\;,
$
$
R_{X_{\alpha}}=R_{\gamma_r}(\frac{\partial\gamma_r}{\partial X_{eff}})^2(\frac{\partial X_{eff}}{\partial X_{\alpha}})^2\;,
$
and
$
R_{X_{a}}=R_{\gamma_r}(\frac{\partial\gamma_r}{\partial X_{eff}})^2(\frac{\partial X_{eff}}{\partial X_{a}})^2
$
with
$
R_{\gamma_r}=\lim_{\gamma_0\tau\to 0}\frac{F_{\gamma_r}(|-\rangle\langle-|,\gamma_0\tau)}{\gamma_0 \tau}=\frac{1}{2(\gamma_r-1)}\;.
$
Since $(\frac{\partial X_{eff}}{\partial X_{\alpha}})^2+(\frac{\partial X_{eff}}{\partial X_{a}})^2=1$, we have
\begin{equation}
R_{X_{eff}}=R_{X_{\alpha}}+R_{X_{a}}\;.
\end{equation}
However, the existence of inherent acceleration does not always degrade the encoding rate of intrinsic temperature. In Fig.~(\ref{F2}), we plot $R_{X_{\alpha}}$ as function of $X_a$ with intrinsic temperature $X_{\alpha}=0.17, 0.1, 0.08$ respectively.
Results show that, in small intrinsic temperature cases, with the increase of Unruh temperature, the encoding rate of the intrinsic temperature increase to its maximum value at first, and then decreases to zero. Therefore, the exist of non-zero inherent acceleration does not always decrease the encoding rate of the intrinsic temperature, and there exists an optimal inherent acceleration corresponding to the largest encoding rate of the intrinsic temperature.
Using the 26\,MHz transition between highly excited Rydberg states (e.g., $n\sim630$--$640$ in carbon or cesium~\cite{1}), we estimate the dimensionless parameter $X_{\alpha}=1/(2\pi\alpha\omega_0)$ to be approximately $7.2\times10^{-15}$ in natural units, which is small enough, and the inherent acceleration may have positive effects on the encoding rate of intrinsic temperature. Therefore, for $X_{\alpha}=10^{-15}$, the largest encoding rate of $X_{\alpha}$ is calculated as $R_{X_{\alpha}}=1.163\times10^{-28}$ when $X_a=0.163$. To estimate $X_{\alpha}$ with uncertainty in the same magnitude of $X_{\alpha}$, the required total probe time becomes $\gamma_0T_{X_{\alpha}}=\frac{1}{(\Delta X_{\alpha})^2R_{X_{\alpha}}}\sim10^{58}$, which is extremely large and is hard to realize in experiment. To reduce the required total probe time, atom with much smaller transition frequency should be used.
\section{Probing Unruh effect in de sitter space-time}
\begin{figure}[htbp]
\centering
\includegraphics[height=2.1in,width=3.1in]{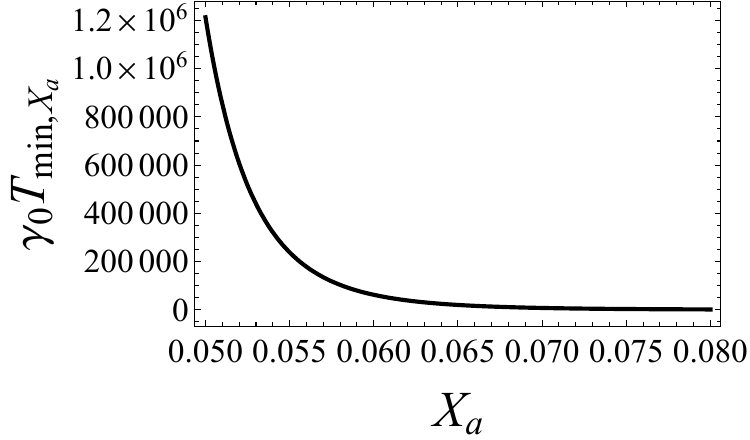}
\includegraphics[height=2.1in,width=3.1in]{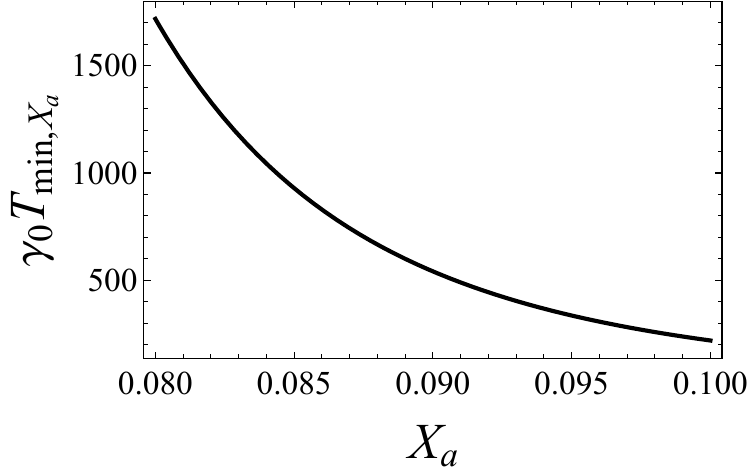}
\caption{ $\gamma_0T_{min,X_a}$ as function of separation $X_a$.
}\label{F3}
\end{figure}

If we put the static probes in positions separated from the free falling observer, the inherent acceleration occurs. Near the cosmological horizon, $X_a\gg X_{\alpha}$, the Unruh effect dominates the thermal effects.
To test the Unruh effect induced by the inherent acceleration, $\Delta X_a=\sqrt{Var (X_a)}$ should be smaller than $X_a$. We let $\Delta X_a\sim X_a$, the minimal total probe time becomes
\begin{equation}
\gamma_0T_{min,X_a}\sim \frac{1}{X_a^2\mathcal{R}_{X_a}}\;.
\end{equation}
Here $X_a$ can be expressed as
\begin{eqnarray}
X_a=X_{\alpha}\frac{r}{\alpha}\big(1-\frac{r^2}{\alpha^2}\big)^{-1/2}.\label{a}
\end{eqnarray}
The magnitude of $X_{\alpha}$ affects the magnitude of minimal total probe time. To show such influence, we choose small transition frequency atom to set $X_{\alpha}=1$.
In Fig.~(\ref{F3}), we plot $\gamma_0T_{min,X_a}$ as function of dimensionless separation $X_a$.
Results show that, with the increase of $X_a$, the required dimensionless total probe time $\gamma_0T_{min,X_a}$ decreases fast. When the $X_a\rightarrow0.1$, $\gamma_0T_{min,X_a}$ reduces to the magnitude of $10^2$, which is feasible in experiment. If we choose $\gamma_0\tau\sim0.1$, which makes the optimal condition $\gamma_0\tau\rightarrow0$ almost fulfilled, the number of probes becomes in magnitude of $10^3$. Such number is also feasible in experiment. Setting $X_{\alpha}\sim10^{-15}$, the separation between the probes and the horizon along radius direction is about $\frac{\Delta r}{\alpha}=5\times 10^{-29}$ according to Eq.~(\ref{a}).

\section{Conclusion}

In summary, we have investigated the thermalization behavior of two-level probes in de Sitter spacetime from a quantum-metrological perspective. By considering a fixed total probe time, we quantify the temperature encoding rate in terms of the Fisher information accumulated per unit total probe time. Our results show that the encoding rate of the effective temperature is equal to the sum of the encoding rates of the intrinsic temperature and the Unruh temperature associated with the probe's inherent acceleration. This additive relation provides a metrological characterization of the effective temperature in terms of its intrinsic and Unruh contributions.

The effect of the position-dependent inherent acceleration on the encoding of the intrinsic temperature is non-monotonic, and an optimal inherent acceleration exists for maximizing the corresponding encoding rate. We have also investigated the encoding rate of the Unruh temperature and the minimum total probe time required to estimate it with sufficient precision as functions of Unruh temperature near the cosmological horizon. The results show that the required total probe time and the corresponding number of probes remain experimentally feasible when $X_a=0.1$.

\begin{acknowledgments}
This work was supported by the National Natural Science Foundation of China under Grants No. 12165003, the special funding of talent program in Guizhou province[GCC[2023]005].
\end{acknowledgments}


\end{document}